\documentclass[12pt]{iopart}
\usepackage{iopams}
\usepackage{graphicx}
\usepackage{subfigure}
\usepackage{CJK}  
\begin{document}
\begin{CJK*}{UTF8}{gbsn} 
\title{Two-photon assisted clock comparison to picosecond precision}
\author{Shi-Wei Zhang (张世伟)$^{1,2,3}$, Jia-Zheng Song (宋家争)$^{1,3}$, Yin-Ping Yao (姚银萍)$^1$, Ren-Gang Wan (万仁刚)$^1$ and Tong-Yi Zhang (张同意)$^1$}
\address{$^1$ State Key Laboratory of Transient Optics and Photonics,
Xi'an Institute of Optics and Precision Mechanics, Chinese Academy of Science,
Xi'an 710119, China}
\address{$^2$ School of Science, Xi'an Jiaotong University, Xi'an 710049, China}
\address{$^3$ University of Chinese Academy of Sciences, Beijing 100049, China}
\ead{zsw@opt.ac.cn}

\begin{abstract}
We have experimentally demonstrated a clock comparison scheme utilizing
time-correlated photon pairs generated from the spontaneous parametric down conversion process of a laser pumped beta-barium borate crystal. The coincidence of two-photon events are analyzed by the cross correlation of the two time stamp sequences. Combining the coarse and fine part of the time differences at different resolutions, a 64 ps precision for clock synchronization has been realized. We also investigate the effects of hardware devices used in the system on the precision of clock comparison. The results indicate that the detector's time jitter and the background noise will degrade the system performance. With this method, comparison and synchronization of two remote clocks could be implemented with a precision at the level of a few tens of picoseconds.
\end{abstract}

\maketitle

\section{Introduction}
High accurate clocks  comparison and synchronization are important for
fundamental physics research and practical applications, such as, Global
Navigation Satellite Systems (GNSSs), power transmission grid, telecommunication, distributed network, etc. There are several modern protocols to synchronize remote clocks, such as network time protocol (NTP), precision time protocol (PTP), and GNSSs\,\cite{eidson2006measurement}. NTP is a method to synchronize clocks by means of message passing over the internet. Its nominal accuracy is in the low tens of milliseconds on wide area networks to sub-milliseconds on a local area
network. For the PTP described in IEEE 1588\,\cite{4579760}, it is
possible to synchronize distributed clocks with an accuracy of less than 1
microsecond via Ethernet networks. In the widely used GNSS scheme, the
two-way satellite time and frequency transfer method\,\cite
{0026-1394-45-2-008}, which takes advantage of exchanging modulated signals
between two sites to cancel most effects that impact on the accuracy, is at
the level of one nanosecond\,\cite{0026-1394-45-2-008, 5422517}.

Recently, some new methods both in theory and experiment have been reported
to improve the accuracy of remote clock synchronization to the level of few
picoseconds. One is the time transfer through optical fibers, which
can reach the accuracy of 100 ps due to the wider bandwidth of the
transmission\,\cite{6533649, 6533650, 0957-0233-21-7-075302,ars-9-1-2011,
0026-1394-49-6-772}. Another way is the time transfer by laser link (T2L2),
by using the propagation of light pulses between satellite and ground clocks
or between remote clocks on the earth\,\cite{0026-1394-35-3-3}. The expected
performance of T2L2 is in the 100 ps range for accuracy, with an ultimate
time stability about 1 ps over 1,000 s and 10 ps over 1 day\,\cite
{4053864, 6533652}. With the help of quantum entanglement, several quantum clock synchronization schemes have been proposed in theory\,\cite{PhysRevLett.85.2006, PhysRevLett.85.2010, PhysRevA.65.022309,PhysRevA.72.042301}, they provide a new way to synchronize remote clocks, and the accuracy can even reach a level of the standard quantum limit. Experimental demonstration of quantum clock synchronization using entangled photon pairs has been carried out by Valencia et al.\,\cite{Appl.Phys.Lett.85.2655}. This method relied on the measurement of the
second-order time-correlation function of entangled photons, and can achieve a
resolution of picoseconds. Ho et al. provided an algorithm to detect the time
and frequency differences of independent clocks based on the remote
coincidence identification of time-correlated photon pairs\,\cite
{1367-2630-11-4-045011}. Using the algorithm, remote clocks can be
synchronized without dedicated coincidence hardware or very stable reference
clocks with nanosecond precision.

In this paper, we present a remote clock comparison experiment using
time-correlated photon pairs generated from the spontaneous parametric down
conversion (SPDC) process of a laser pumped beta-barium borate (BBO)
crystal. Using the cross correlation method, we have identified the
time-correlated photon pairs events that detected by two remote
clocks. The time difference of the two clocks is obtained via the cross correlation and peak searching process at a 64 ps resolution, and the standard deviation of the time difference is 55.92 ps. This scheme can be used in quantum communication and quantum position systems to improve the precision of clock synchronization.

\section{Cross correlation method for clock comparison}

The process of comparing remote clocks with time-correlated photon pairs is
similar to the problem of time delay estimation\,\cite{1162830,1457990}. The
arrival times of single photons are recorded as time stamps, and it can be
mathematically modeled as, 
\numparts
\begin{eqnarray}
A(t) &=& s_{1}(t)+n_{1}(t)\\
B(t) &=& s_{2}(t)+n_{2}(t)
\end{eqnarray}
\endnumparts
where $s_{1,2}(t)$ are the arrival time sequences of the photons to detectors, and $n_{1,2}(t)$ represent the erroneous time stamps
caused by detector's dark counts and background light. The signals $s_{1,2}$ are assumed to be uncorrelated with noises $n_{1,2}$, and
the noises $n_{1,2}$ are also uncorrelated with each other.

For the comparison of two remote clocks, we calculate the cross correlation of two
time sequences $A(t)$ and $B(t)$. By a linear searching process, the time
difference of clocks can be found in the peak position of cross correlation
function, i.e.
\begin{equation}  \label{eqn:max}
T_{diff} = \mathop{arg}_{t} \max[R_{c}(t)]
\end{equation}
where $R_{c}$ is the cross correlation of $A(t)$ and $B(t)$, and can be
calculated via fast Fourier transformations (FFT) and their inverse\,\cite
{1367-2630-11-4-045011}, 
\begin{equation}  \label{eqn:cc}
R_{c} = E[A(t)B(t)] = \mathcal{F}^{-1}[\mathcal{F}^{*}[A] \cdot \mathcal{F}
[B]]
\end{equation}
where $E$ denotes the expectation; $\mathcal{F},\,\mathcal{F}^{-1}$
represent FFT and it's inverse; superscript * indicates complex
conjugate operation.

\section{Clock comparison experiment}

\label{sec:exp_result}

\begin{figure}
\begin{center}
\includegraphics[width=0.8\columnwidth]{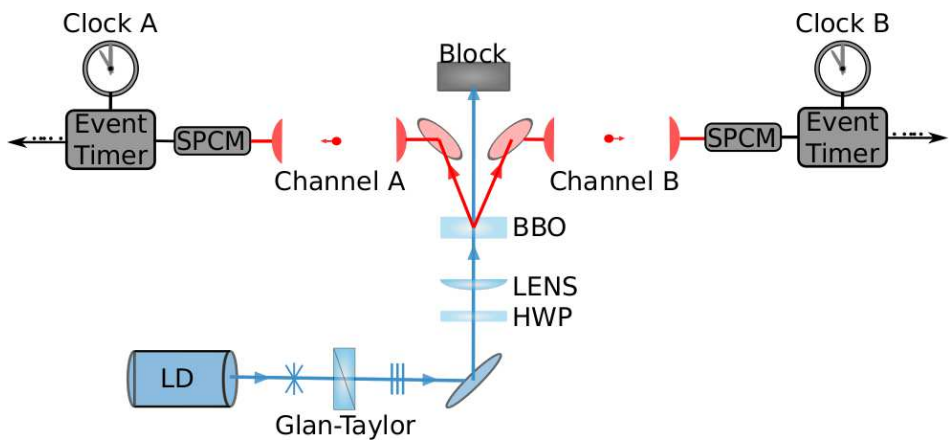}
\caption{The experimental setup of remote clocks comparison scheme.} \label{fig:clockcomp}
\end{center}
\end{figure}

\subsection{Experimental setup}
\label{sec:setup} 
The schematic of the clock comparison experiment is given in \fref{fig:clockcomp}. A 403 nm laser diode with 30 mW power impinges on a
type-II non-collinear cutting 2 mm long BBO crystal to generate two
down-converted photons with degenerated wavelength centered at 806 nm. The
two single photons are then distributed via channel A and channel B to two
remote sites, where there are two clocks (clock A and clock B) need to be synchronized. After flying in free space and collected by optical systems, single photons are
detected by Si avalanche photo diodes (PerkinElmer SPCM-AQRH-16) with
response-time jitter about 350 ps\,\cite{PhysRevLett.92.113601}. The arrival times of single photons at the detectors are recorded by two
event timers (A033-ET). The event timer is connected with a rubidium
clock (SRS 625 with the frequency stability $3.16 \times 10^{-11}$ in 10 seconds) and locked to the rubidium clock's 10 MHz oscillator. When the detector receives a single photon, the event timer generates
an output signal of the arrival time of the single photon. So, during the acquisition time, we get two sequences of arrival time stamps $\{A(t)\}$
and $\{B(t)\}$.

\begin{figure}
\begin{center}
\includegraphics[width=0.8\columnwidth]{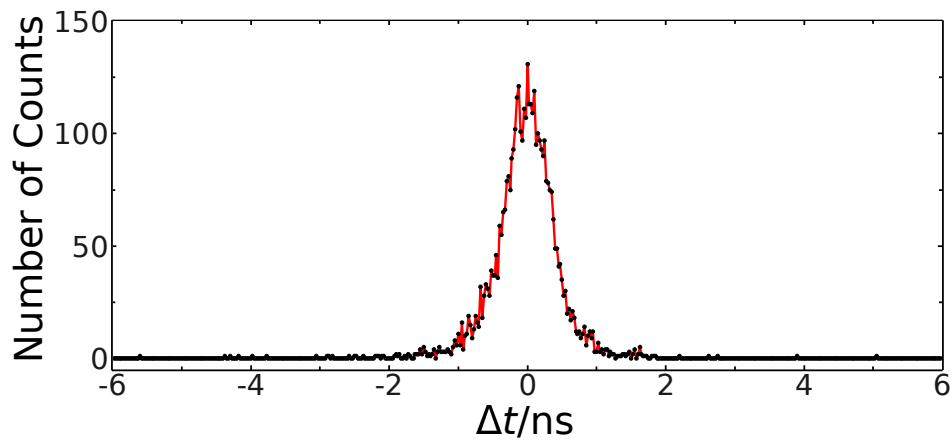}
\caption{The coincidence time distribution of two-photon. The horizontal axis $\Delta t$ represents the relative delay of two
single photons, and the vertical axis is the number of coincidence counts in
corresponding relative delay.}
\label{fig:mca}
\end{center}
\end{figure}

To characterize the time-correlation of down-converted photons, we
employ the hardware coincidence method to measure the coincidence time
distribution. In this method, the hardware coincidence equipment consists of
a nanosecond delay (ORTEC 425A), a time-to-amplitude converter (TAC), and a
multi-channel analyzer (MCA). One of the output signal from the two single
photon detectors is used as the start signal of TAC, and the other one as the stop signal after passing through a nanosecond delay, the coincidence time distribution of the two-photon is recorded in MCA by
measuring the the pulse height distribution of the output of TAC. The result
is shown in \fref{fig:mca}, in which the time axis represents the relative delay of two photons, the full-width at half-maximum of the coincidence time distribution peak is 0.698 ns. It's a combine effect of detector jitter,
electric delay in TAC, and the bin size of MCA, but mainly due to the detector's response-time jitter. A bandpass interference filter with 3 nm
bandwidth and $50\%$ transmissivity is used to block the background light. During a 10 seconds sampling period, 4207 coincidence counts are recorded. The average coincidence detection rate is about 420 counts per second to ensure that there are sufficient number of two-photon events to find the time difference of remote clocks.

\subsection{Time difference calculation}
\label{sec:time_diff_cal} 
Generally, the common method for finding the time difference is to
calculate the cross correlation of two signals and to select the peak of in
cross correlation result as the estimate of the time difference $T_{diff}$.

In the clock comparison process, if we directly calculate the cross
correlation, the size of FFT arrays will exceed $10^{10}$ (for time resolution
of 10 ps and 0.1 s acquisition time). It's a time-consuming
process and also impractical. To solve this problem, we use the method
proposed by Ho et al.\,\cite{1367-2630-11-4-045011}, to calculate the coarse
and fine parts of $T_{diff}$ separately in a moderate size. Then,
combining the results of the two parts, a high resolution result can be
obtained. In this method, the time stamp sequences are converted into
discrete arrays with a time resolution $\delta t$, 
\begin{equation}
A_{k} = \sum_{i} \delta_{k, \lfloor (t_{i} / \delta t) \;\mbox{mod} \;N \rfloor}
\quad k = 0, \cdots, N-1
\end{equation}
where $N$ is the size of discrete arrays, and the same for $B_{k}$. Then,
substitute $A_{k}$ and $B_{k}$ into the equations \eref{eqn:max} and \eref
{eqn:cc}, we can obtain the peak position corresponding to a resolution $\delta t$ and modulo $N\delta t$.

The time difference of two time stamp sequences may be very long that a large size of the discrete array is needed. In practice, we adjust the two sequences and make them align left. Firstly,
the difference of the first time stamps of each sequence, named as $
\Delta T_{1}$, is calculated by a subtract operation. Based on $\Delta T_{1}$, the two sequences are adjusted to make the
first element of each sequence identical. Then, the relative offset of two
revised sequences $\Delta T_{2}$ is calculated by the cross correlation method. Finally, by simply adding the relative offset $\Delta T_{2}$ to the first time stamp difference $\Delta T_{1}$ , the time difference of two time stamp sequences can be expressed as $\Delta T = \Delta T_{1} + \Delta T_{2}$. 

In the cross correlation calculation, the noise events may lead to a high
background which seriously smooth the correlation peak. In the worse case, it maybe give a false result. Generally, the noise events can be remit by using the prefilter method in generalized correlator\,\cite{1163269}. We use a simple method to get rid of the obvious noise events. At first, check the adjacent time stamps, if they are from the
same detector, then drop the previous one and compare with the next one,
continue until the adjacent events are from different detectors. Secondly, set a threshold $\Delta t_{th}$, and remove the adjacent pairs with time
difference exceed the threshold. With this method, most of the obvious
uncorrelated events can be eliminated and only the events with the most
probability to be similar to the true two-photon pairs maintained.

\subsection{Clock comparison result}

\label{sec:comp_result} 
For a efficiently cross correlation calculation, we define the size of cross correlation array $N=2^{23}$, and set the coarse resolution as $2^{15}$\,ps. The
corresponding acquisition time is $2^{38}$\,ps to make sure that there are enough photon pairs events in the subset of the time stamp sequence. One subset is extracted during the acquisition time and there are 3097 time stamps from
sequence A, and 4094 from sequence B. The first time stamps difference $\Delta T_{1}$ of the two subset
sequences is $1716\,789\,048\,793$\,ps. Then, the two sequences are revised, and the relative offset $\Delta T_{2}$ is calculated using the method in \sref{sec:time_diff_cal} with different resolutions in $2^{15}$\,ps, $
2^{10}$\,ps, $2^{6}$\,ps, and $2^{5}$\,ps, respectively.

\begin{figure}
\begin{center}
\subfigure[\label{peak_15}$Res=2^{15}, Pos =592$]
{\includegraphics[width=0.4\columnwidth]{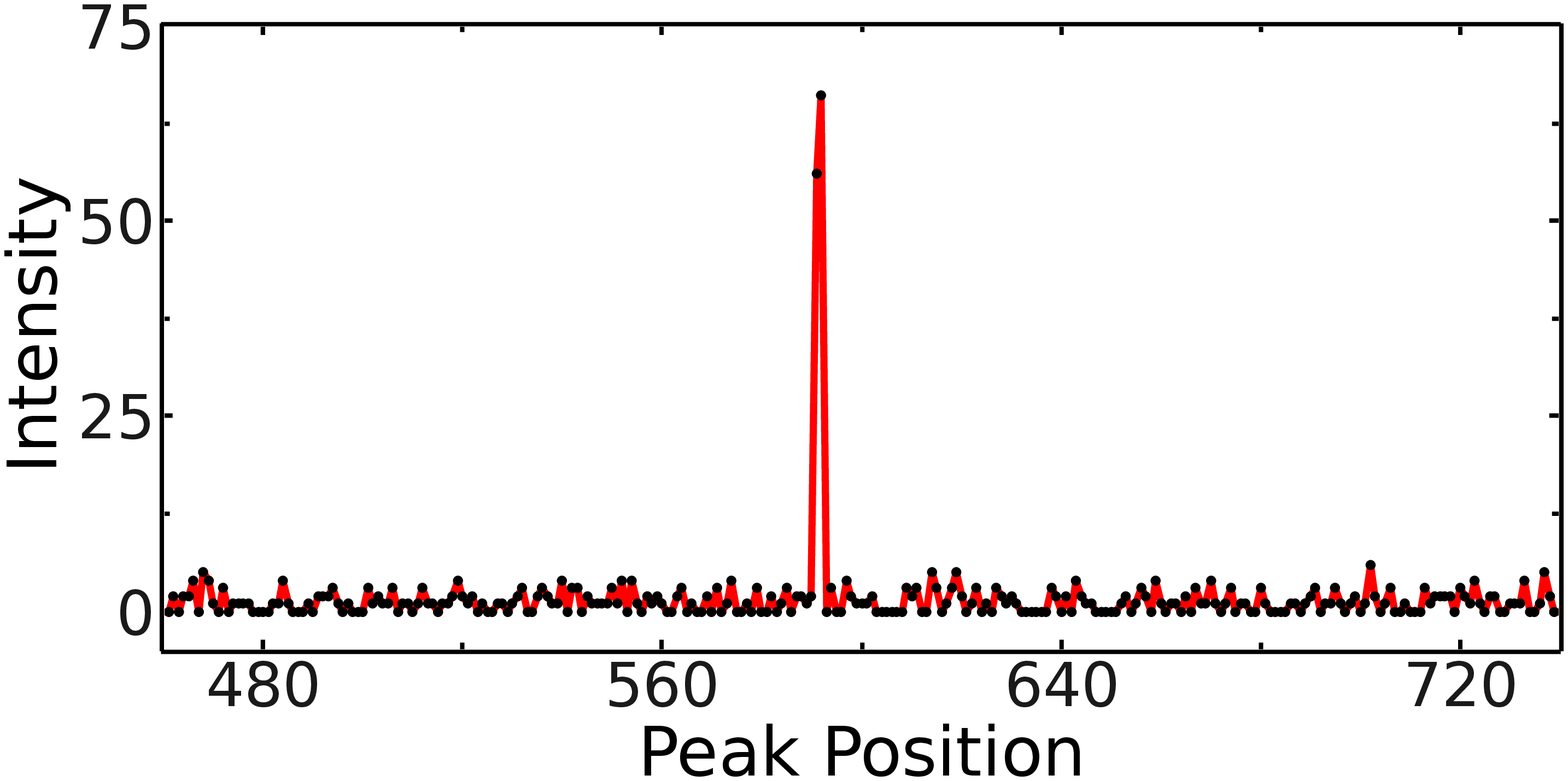}}
\hspace{6pt}
\subfigure[\label{peak_10}$Res=2^{10}, Pos =18\,929$]
{\includegraphics[width=0.4\columnwidth]{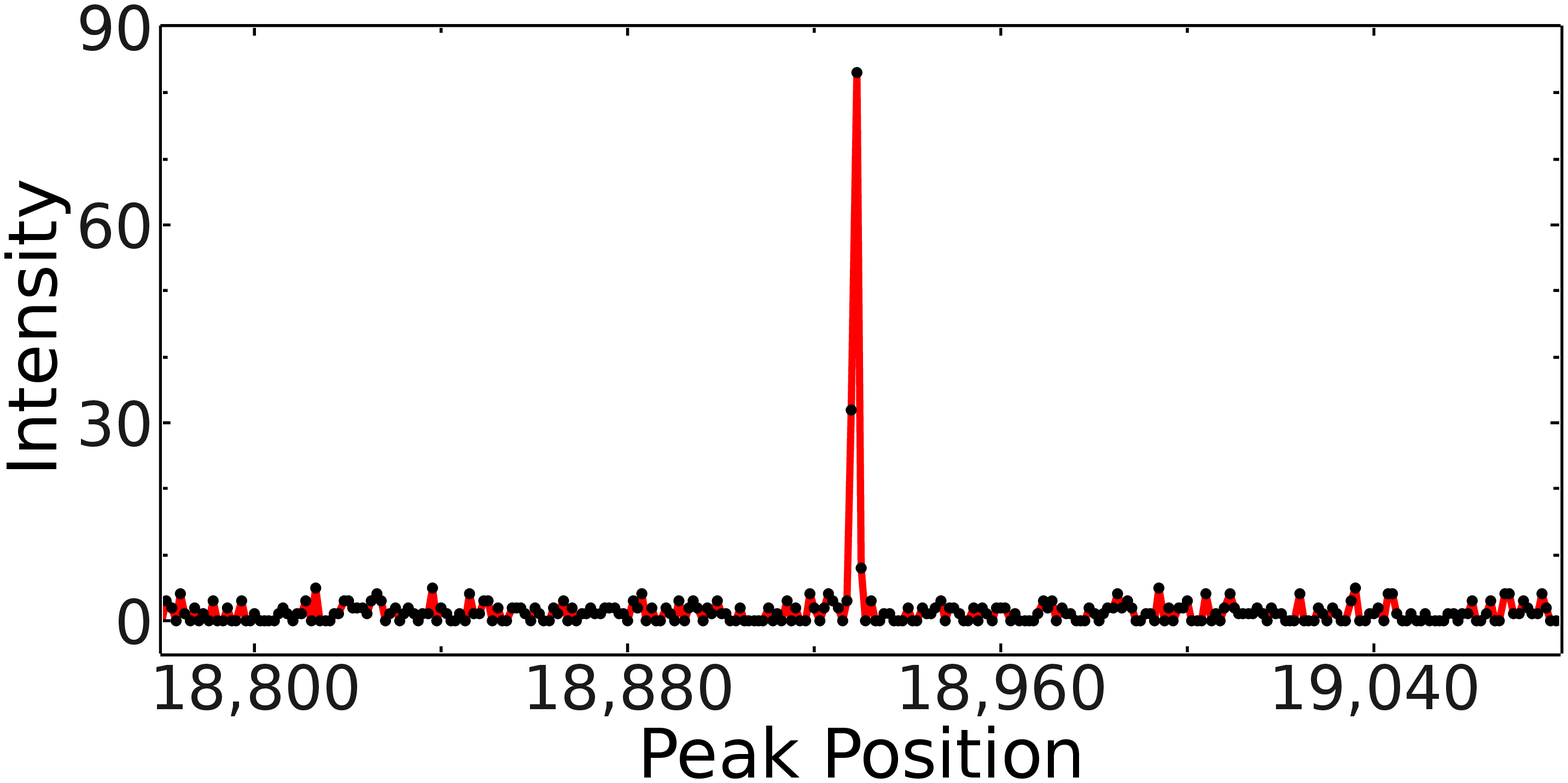}}
\subfigure[\label{peak_6}$Res=2^{6}, Pos =30\,2861$]
{\includegraphics[width=0.4\columnwidth]{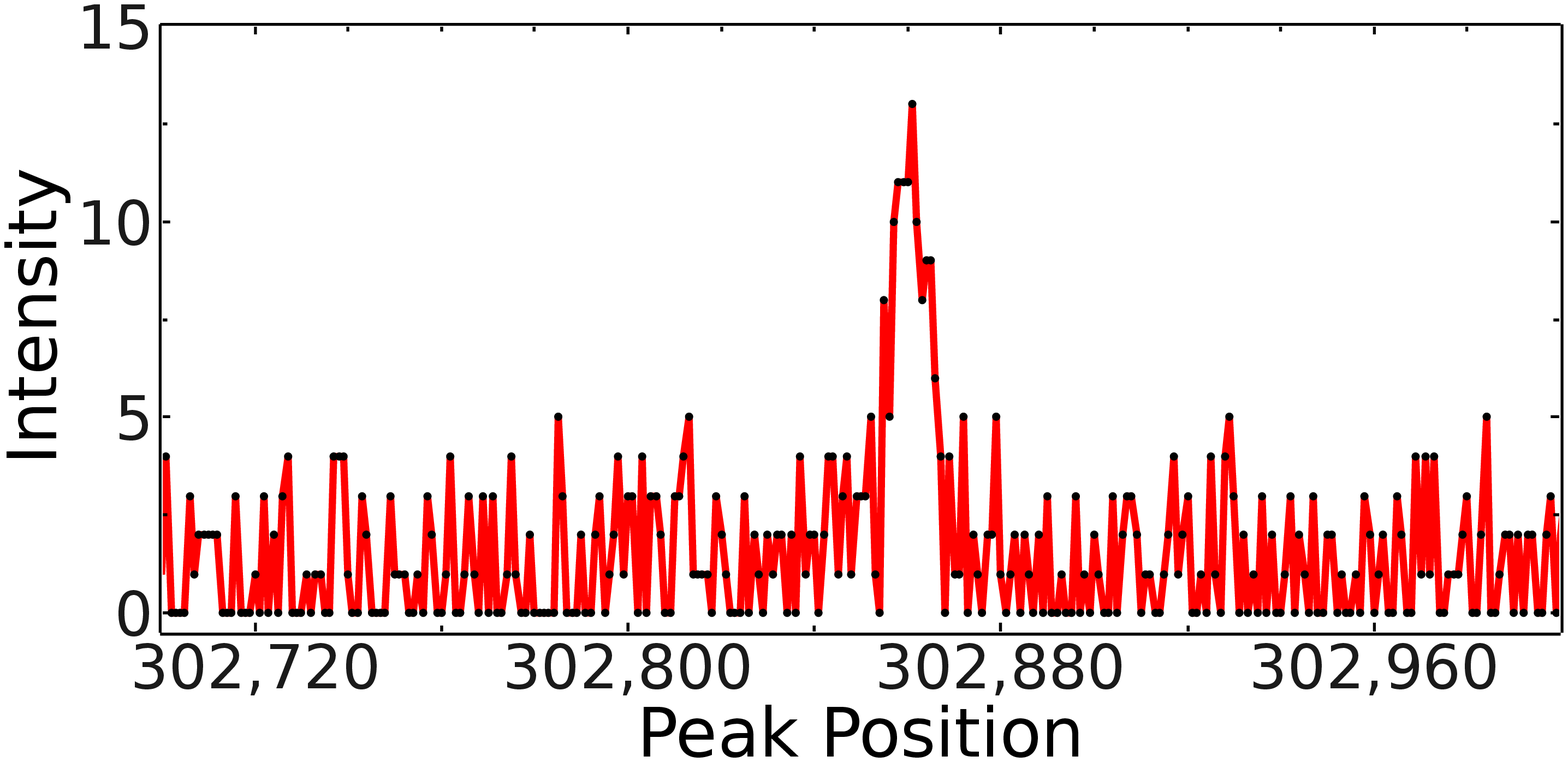}}
\hspace{6pt}
\subfigure[\label{peak_5}$Res=2^{5}$]
{\includegraphics[width=0.4\columnwidth]{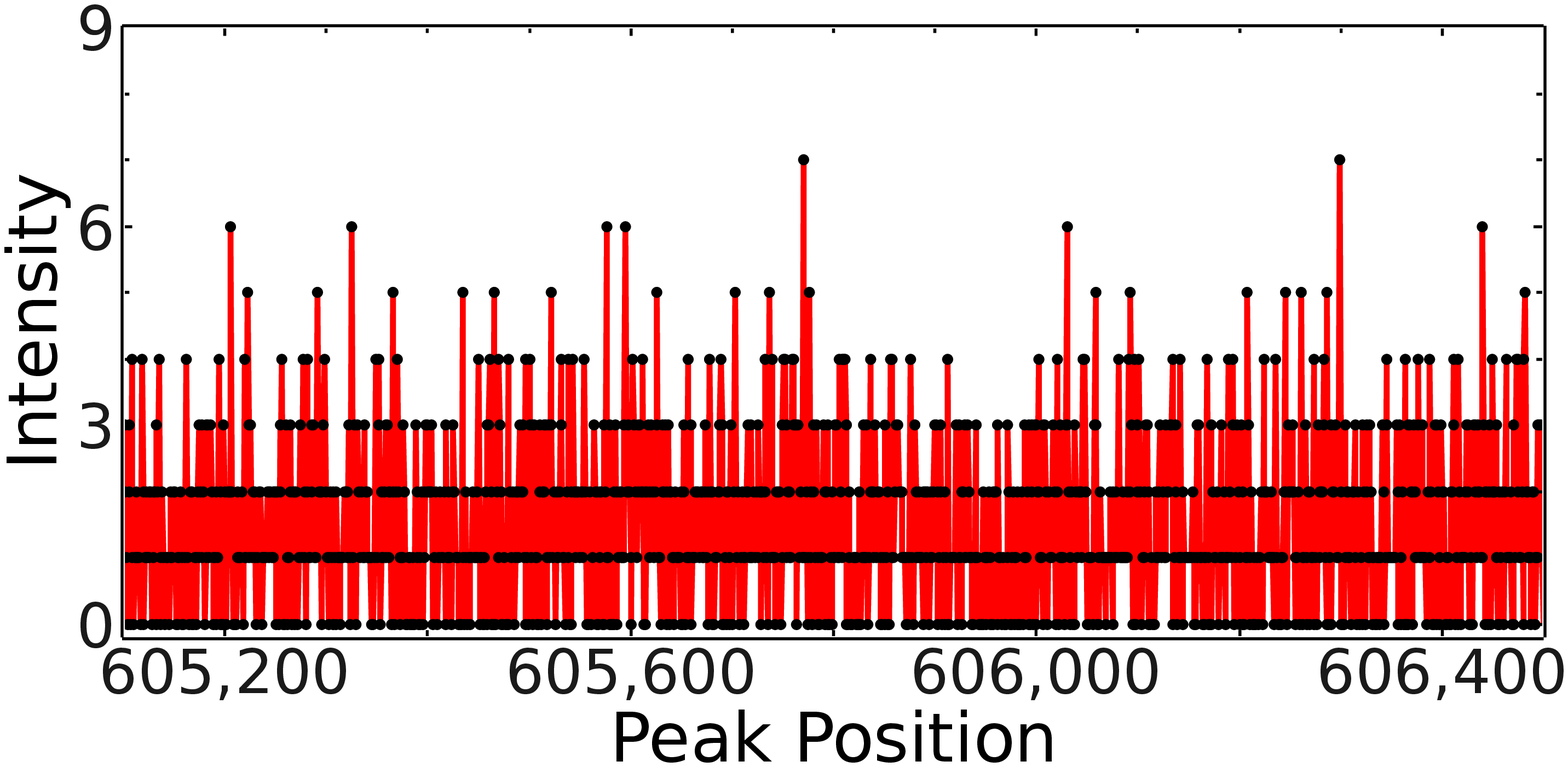}}
\caption{The cross correlation results of two revised sequences at different resolutions. $Res$ represents the resolution, and $Pos$ indicates the peak position in the array.}
\label{fig:peak}
\end{center}
\end{figure}

The cross correlation results of the two revised sequences are show in \fref{fig:peak}. The peaks in \fref{peak_15} at resolution $
2^{15}$\,ps, \fref{peak_10} at resolution $2^{10}$\,ps, and \fref{peak_6} at resolution $2^{6}$\,ps are sharp enough to be identified with sufficient
significance. Consider the situation at resolution $2^{5}$\,ps or below,
as shown in \fref{peak_5}, it's hard to uniquely
identify an obvious signal peak. In such case, the cross correlation of the two-photon events is submerged in noise that mainly caused by the jitter of detectors and the background light. 

The clock comparison results of the selected subset time stamp sequences are shown in \tref{tab:result}. The time difference at the coarse resolution $
2^{15}$\,ps is directly calculated and a 1716\,808\,447\,449\,ps difference is
obtained. The time differences at fine resolutions $2^{10}$\,ps and $2^{6}$\,ps
are calculated by combining the least significant byte of the fine results
with the most significant byte of the coarse result that calculated at $
2^{15}$\,ps resolution, respectively. Finally, the time difference at $2^{10}$\,ps resolution
is 1716\,808\,432\,089\,ps and 1716\,808\,431\,897\,ps for the resolution at 
$2^{6}$\,ps. 

\begin{table}
\caption{\label{tab:result}The clock comparison results of one subset at different resolutions.}
\begin{indented}
\item[]\begin{tabular}[l]{@{}lccc}
\br
$\Delta T_{1}$[ps] & \multicolumn{3}{c}{$1716\,789\,048\,793$} \\
\mr
$Res$[ps] & $2^{15}$ & $2^{10}$ & $2^{6}$  \\
\mr
$Pos$ & 592 & 18\,929 & 302\,861  \\ 
$\Delta T$[ps] & 1716\,808\,447\,449 & 1716\,808\,432\,089 & 1716\,808\,431\,897  \\
\br
\end{tabular}
\end{indented}
\end{table}

\begin{table}
\caption{\label{tab:peaks}The clock comparison results of 20 subsets at resolution $2^6$\,ps.}
\begin{indented}
\item[]\begin{tabular}[l]{@{}lccc}
\br
$M^{\rm a}$ & $\Delta T^{\rm b}$\,[ps] & $M^{\rm a}$ & $\Delta T^{\rm b}$\,[ps] \\ 
\mr
1 & 1716\,808\,431\,897 & 11 & 1716\,808\,431\,939 \\
2 & 1716\,808\,431\,950 & 12 & 1716\,808\,431\,935 \\
3 & 1716\,808\,431\,978 & 13 & 1716\,808\,431\,919 \\
4 & 1716\,808\,431\,868 & 14 & 1716\,808\,431\,918 \\
5 & 1716\,808\,432\,016 & 15 & 1716\,808\,431\,848 \\
6 & 1716\,808\,431\,938 & 16 & 1716\,808\,431\,825 \\
7 & 1716\,808\,431\,928 & 17 & 1716\,808\,431\,873 \\
8 & 1716\,808\,431\,896 & 18 & 1716\,808\,431\,849 \\
9 & 1716\,808\,431\,965 & 19 & 1716\,808\,431\,807 \\
10 & 1716\,808\,431\,964 & 20 & 1716\,808\,431\,843 \\ 
\br
\end{tabular}
\item[]{$^{\rm a}$\,The column number indicates the $M$th subset.}
\item[]{$^{\rm b}$\,$\Delta T$ is the time difference corresponding to each subset sequences.}
\end{indented}
\end{table}

We successively extract 20 subsets, with the same acquisition time $2^{38}$\,ps, from the time stamp sequences of $\{A(t)\}$ and $\{B(t)\}$. The  time difference of each subset is calculated at the resolution $2^6$\,ps, as shown in \tref{tab:peaks}.
The mean value of the time differences is $\overline{\Delta T} = 1/M \sum_{i = 1}^{M} \Delta T_i = 1716\,808\,431\,907\,ps$, and the standard deviation is calculate by
\begin{equation}
\sigma = \left[ \sum_{i = 1}^{M}(\Delta T_i - \overline{\Delta T})^2 / (M - 1)\right]^{1/2} = 55.92\,ps.
\end{equation}
Finally, the time difference between two remote clocks is estimated to be $1716\,808\,431\,907 \pm 56\,ps$.

\section{System performance analysis}

\begin{figure}
\begin{center}
\includegraphics[width=0.8\columnwidth]{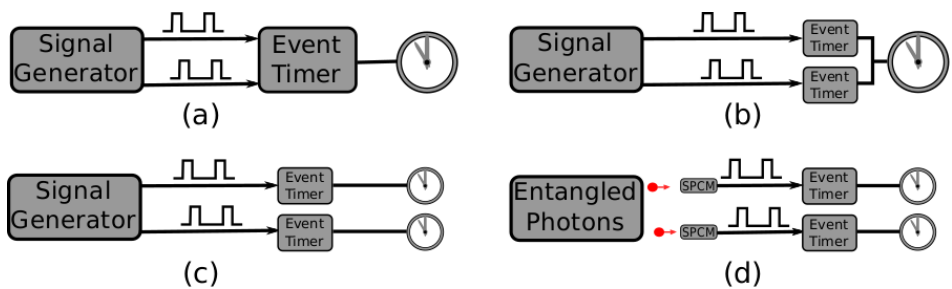}
\caption{Sketch of the schemes I-IV. (a), (b), (c), and (d) correspond to
the schemes I-IV in sequence.} 
\label{fig:compare}
\end{center}
\end{figure}

To analyze the effects of hardware devices, such as the rubidium clocks, event timers, and
single photon detectors on the precision of clock synchronization system, we
propose four different schemes, named as scheme I-IV shown in \fref
{fig:compare}. In schemes I-III, two square
pulse signals with 1MHz frequency and 2\% duty (the pulse width is 20\,ns,
just for simulating the output electric pulse of the single photon detector) from a signal generator
are used as the input signals of the clock comparison system. In scheme I,
two input signals connect with the two channels of one event timer, and the
event timer connects with a clock; for scheme II, each input signal connects
with an event timer, two event timers connect with the same clock; and in
scheme III, each input signal connects with an event timer, but the two event
timers connect with different clocks; the scheme in \sref
{sec:exp_result} is defined as scheme IV, which is the same as scheme III
except that the two input signals are generated from two single photon detectors.

In the schemes I-III, two time stamp sequences are recorded as $\lbrace A\rq{
}(t) \rbrace$ and $\lbrace B'(t) \rbrace$, and each sequence contains
7500 time stamps. The relative time difference of two sequences $\lbrace A
'(t) \rbrace$ and $\lbrace B'(t) \rbrace$ can be directly obtain by
a subtraction operation. The standard deviation of the relative time
difference is calculated to character the performance of system. By
comparing different schemes, the effects of system components can be
analyzed individually.

\begin{figure*}
\begin{center}
\subfigure[\label{fig:5a}]{\includegraphics[width=0.3\columnwidth]{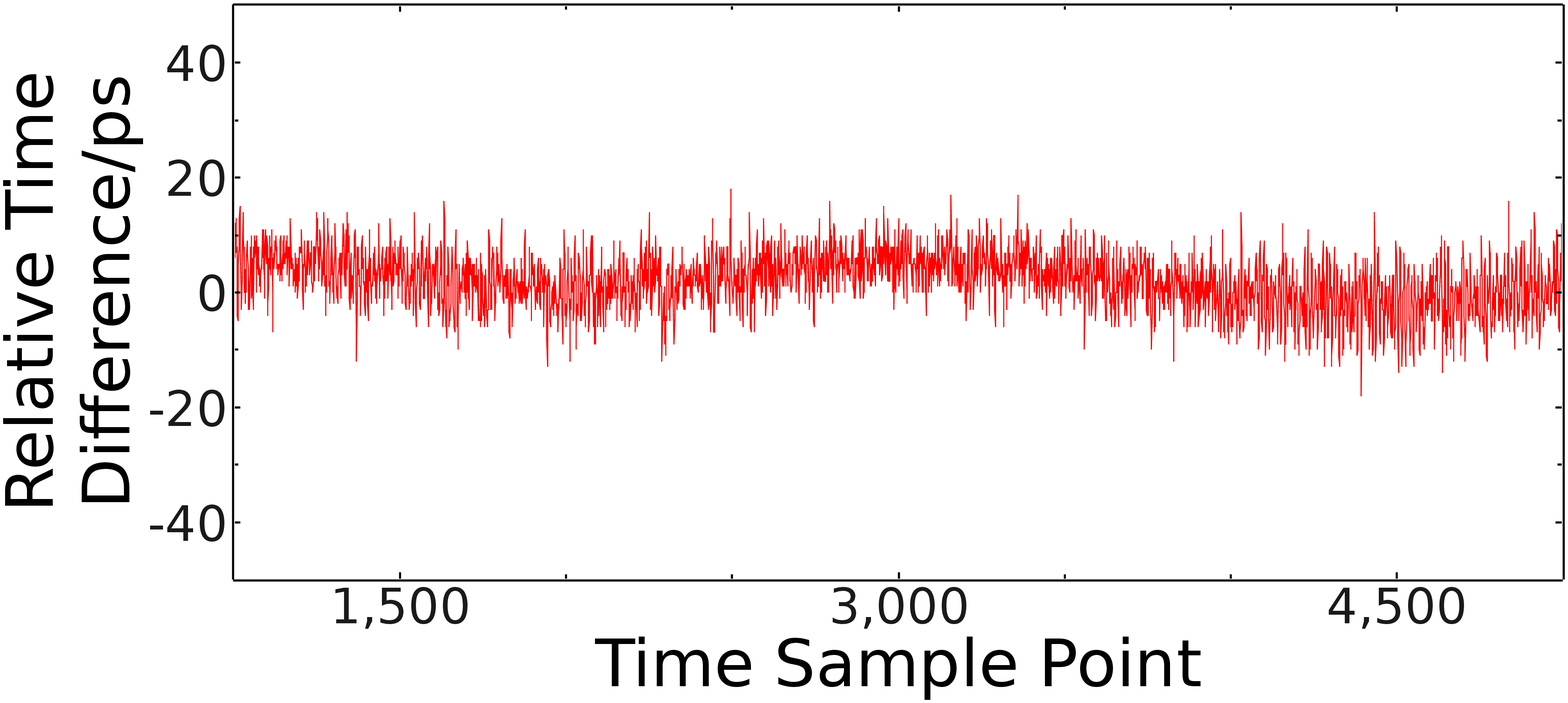}}
\hspace{6pt}
\subfigure[\label{fig:5b}]{\includegraphics[width=0.3\columnwidth]{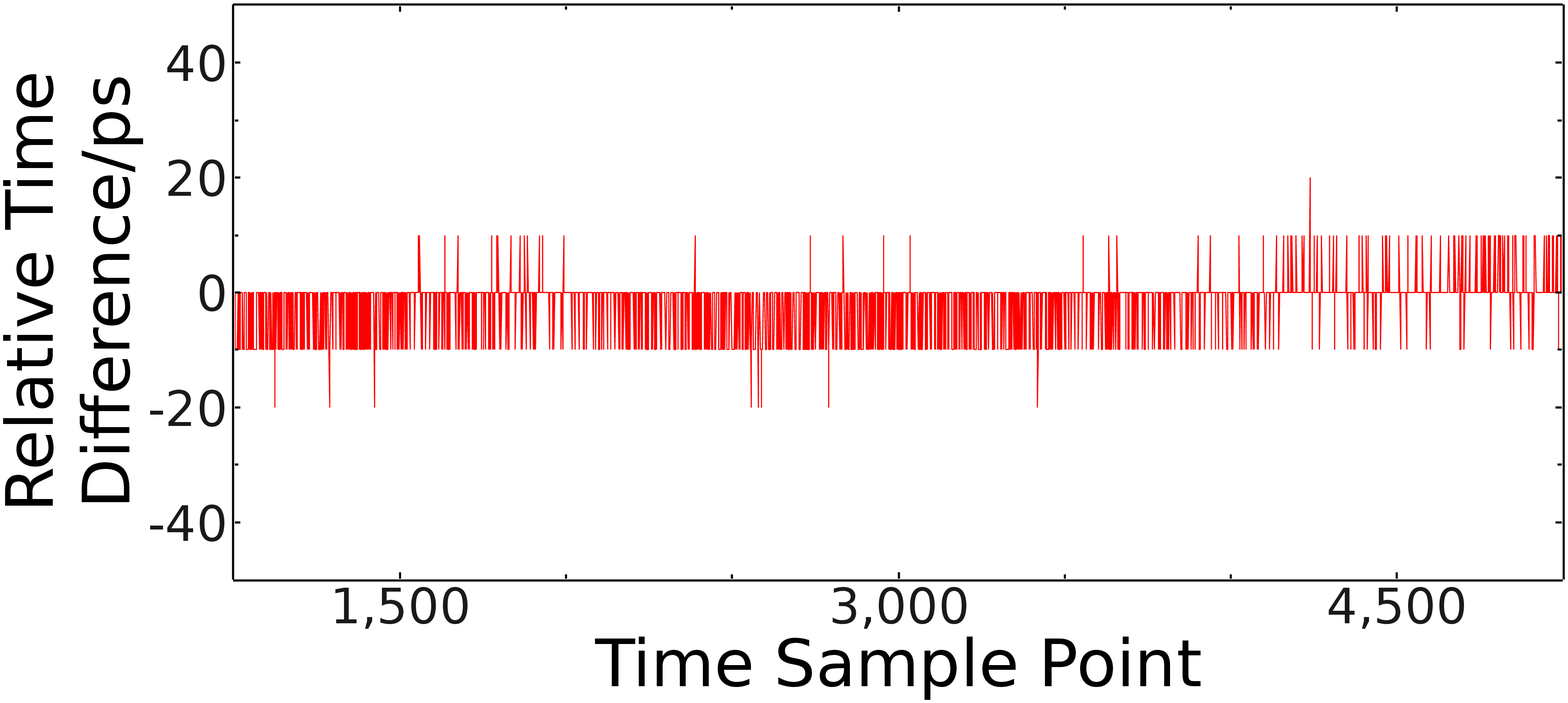}}
\hspace{6pt}
\subfigure[\label{fig:5c}]{\includegraphics[width=0.3\columnwidth]{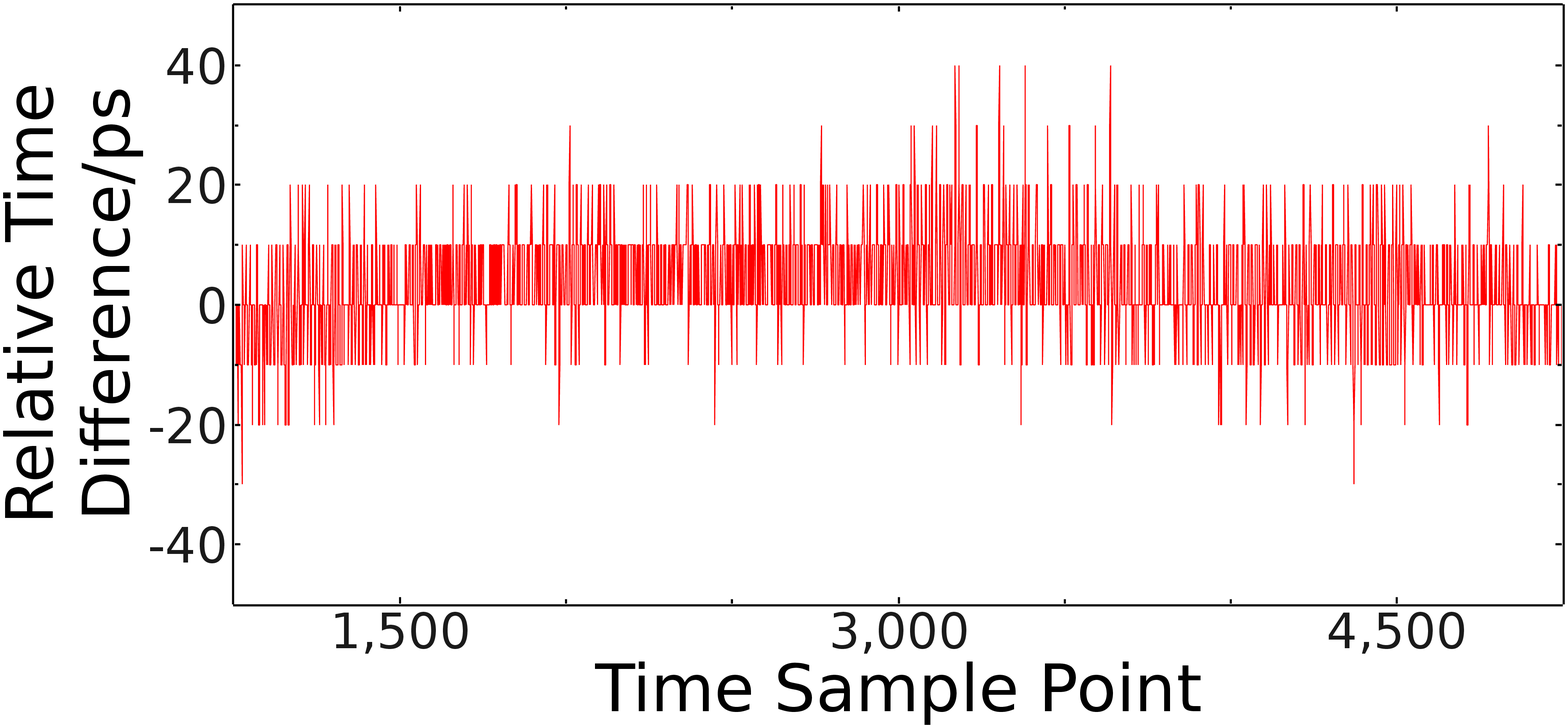}}
\subfigure[\label{fig:5d}]{\includegraphics[width=0.3\columnwidth]{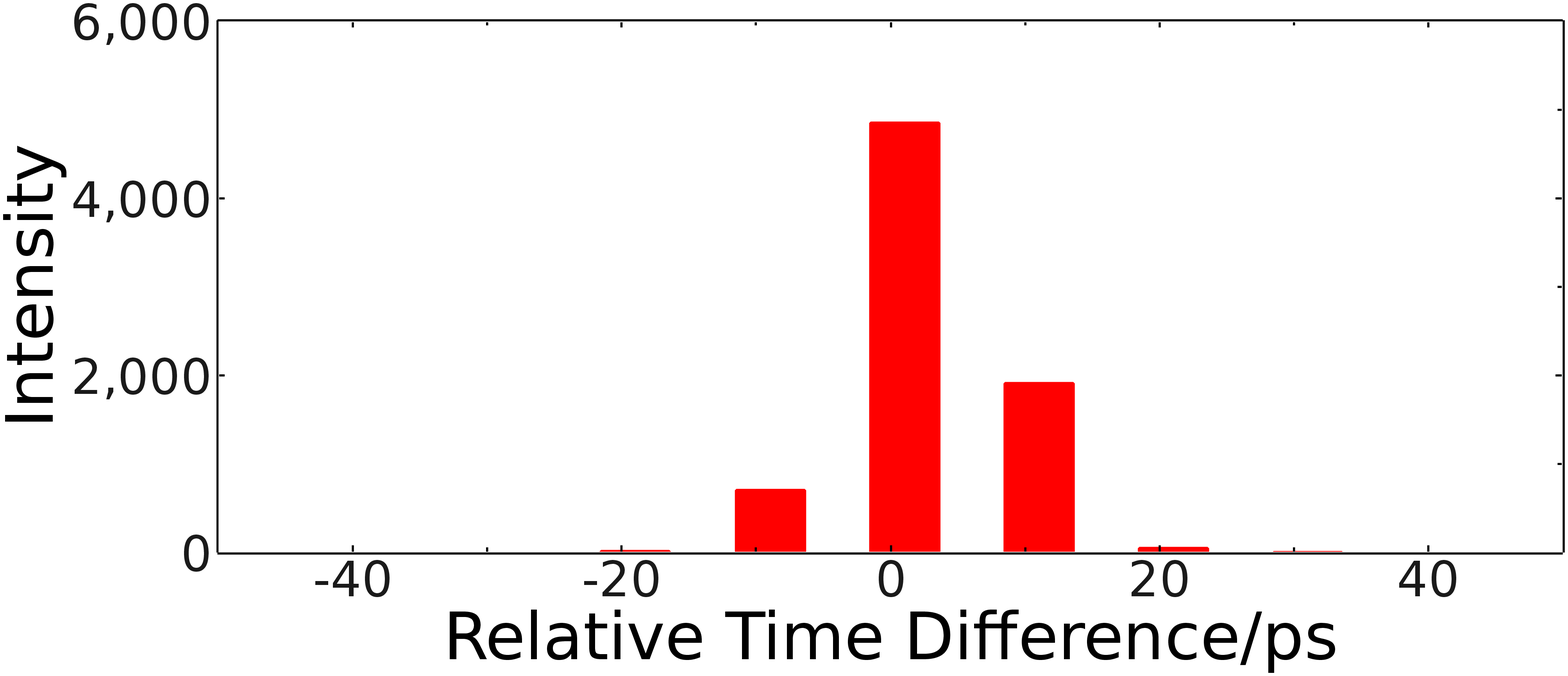}}
\hspace{6pt}
\subfigure[\label{fig:5e}]{\includegraphics[width=0.3\columnwidth]{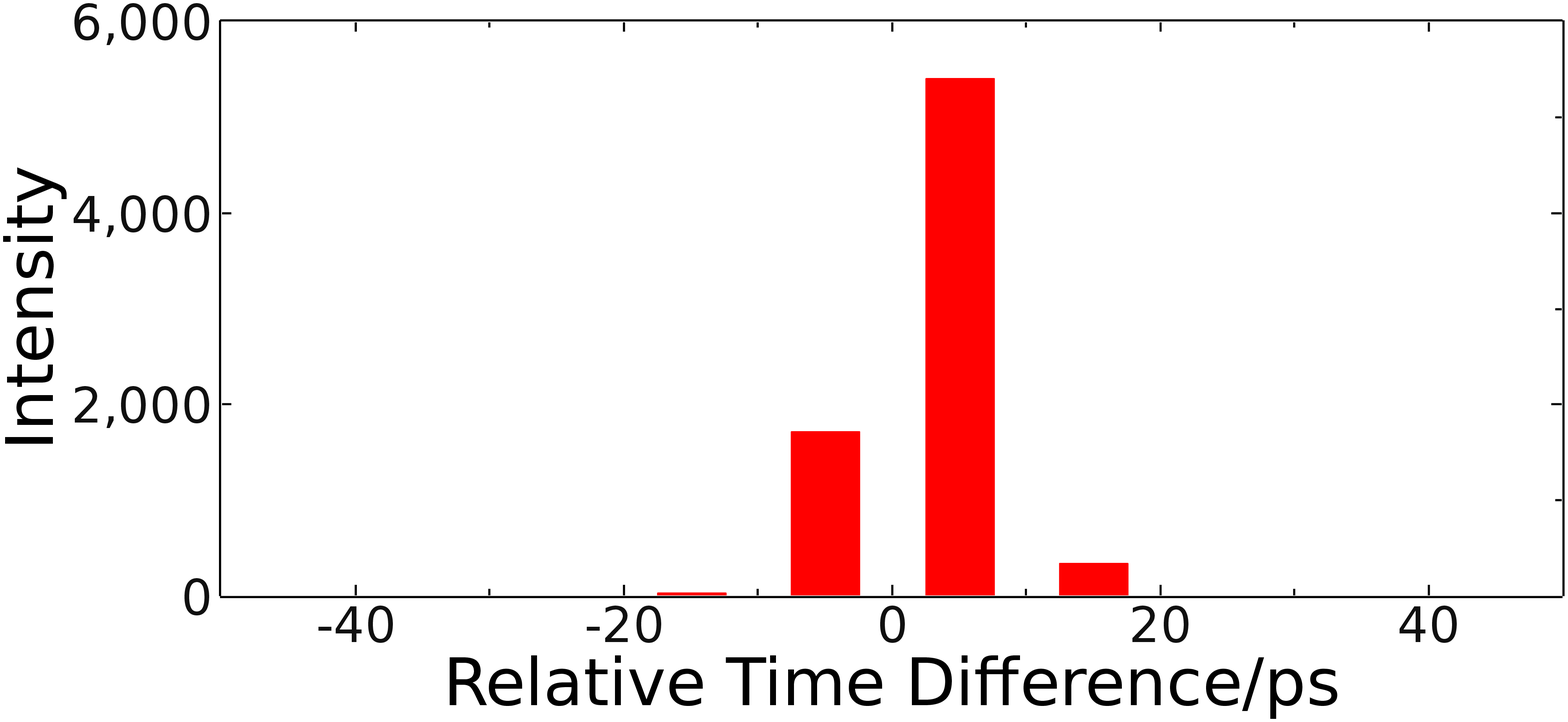}}
\hspace{6pt}
\subfigure[\label{fig:5f}]{\includegraphics[width=0.3\columnwidth]{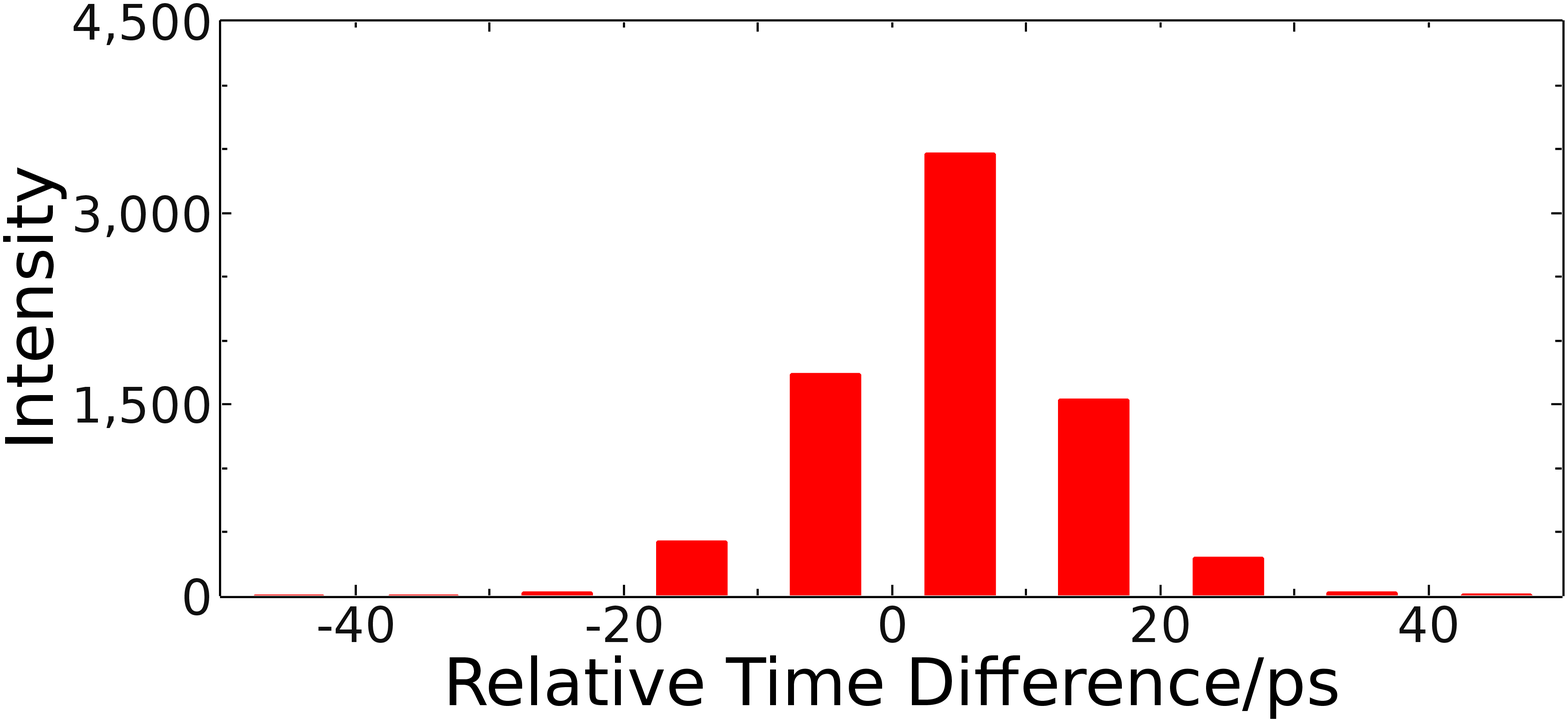}}
\caption{Statistical results for the relative time difference of two time
stamp sequences $\lbrace A'(t) \rbrace$ and $\lbrace B'(t) \rbrace$. \Fref{fig:5a}, \fref{fig:5b} and \fref{fig:5c} illustrate the relative time difference distribution for part of the time sequences. \Fref{fig:5d}, \fref{fig:5e} and \fref{fig:5f} are the histograms of the relative time difference. \Fref{fig:5a} and \fref{fig:5d} in the first column are the results for scheme I, the second and the third columns are the results for scheme II and scheme III, respectively.}
\label{fig:pfm1}
\end{center}
\end{figure*}

At first, the influence of event timers is analyzed by comparing scheme I
with scheme II. We ignore the difference in the two channels of one event timer.
The only difference in these two schemes is the event timer. The standard
deviation of the relative time difference of two input signals in scheme I
is $\sigma_1 = 5.15$\,ps, the distribution and the histogram of the relative
time difference in scheme I is shown in \fref{fig:5a} and \fref{fig:5d}. The
result for scheme II is that $\sigma_2 = 4.67$\,ps, and the statistical
results are shown in \fref{fig:5b} and \fref{fig:5e}. For event timers, the standard deviations of total error in measurement of time intervals between events are calibrated as $\sigma_{ET1} = 2.91$\,ps, $\sigma_{ET2} =3.01$\,ps. So, we can see that there is no obvious difference in the results of scheme I and scheme II, $\sigma_1$ and $\sigma_2$ are almost the same. Therefore, the effect of event
timers on the precision of clock comparison can be ignored when different
equipments are used in remote sites.

Secondly, the effect of clock's frequency offset is investigated by
comparing scheme II and scheme III. If the two clocks used in
experiments have exactly the same frequency, i.e. $\delta u=0$, then
the contribution of two-photon events will all end up in a single time bin
in the cross correlation calculation. While for the case $\delta u\neq 0$,
the correlated events will spread out, and the arrival time stamps of the
single photon streams become $t_{i}^{\prime }=t_{i}\times (1+\delta u)$. This can not only reduce the intensity of the cross correlation
peak, but also increase the width of the peak. In scheme
II the two event timer are connected to the same atomic clock, which corresponds to the case that two clocks with no frequency offset. While in scheme III the two event timer are connected to different atomic clocks, and there may exist some
frequency offset. For scheme III, the standard deviation of the measured relative time difference is $\sigma _{3}=8.99$\,ps, and the statistical results are shown
in \fref{fig:5c} and \fref{fig:5f}. Compared with the results
in scheme II, $\sigma _{3}$ is larger than $\sigma _{2}$. It means that the
offset in clock frequency will decrease the precision of the clock
comparison. As the frequency stability of rubidium clock is
$3.16 \times 10^{-11}$, the time drift during the acquisition time ($2^{38}\,ps \approx 0.275\,s$) is about 8.69\,ps. Thus for the 64\,ps resolution, i.e. a 64\,ps coincidence window for time-correlated photon pair events, the time drift caused by frequency
offset could not affect the cross correlation result. Only in the cases that
the resolution is comparable to the time drift or even small (i.e. time
resolution less than 8.69\,ps), the effects of the time drift caused by frequency offset on clock comparison need to be considered.

Finally, the effect of single photon detector's time jitter is studied by comparing scheme IV with scheme III. The difference of the two schemes is that the
signal from single photon detector has a random time jitter along with background
noise. Similar to the mechanism that the clock's frequency offset affects the cross correlation, the detector's time jitter leads to a random fluctuation of the single photon's arrival time. As shown in \sref{sec:comp_result}, the standard deviation of the time differences in scheme IV at 64\,ps resolution is $\sigma = 55.92$\,ps,  whereas in scheme
III $\sigma _{3}=8.99$\,ps. Therefore, the detector's time jitter and the background noise seriously degrade the accuracy of the clock comparison.

\section{Conclusion}

In conclusion, a 64\,ps precision for clock comparison has been realized using the time-correlated two-photon pairs. The time
difference of two remote clocks is calculated by cross correlation method and
the peak searching process after eliminating the obvious noise events. The influences of the event timer, the clock's frequency
offset, and the single photon detector's time jitter are analyzed by
comparing four schemes with different hardware devices in the clock comparison system. The results show
that the main influence is the detector's time jitter and background noise, the clock's
frequency offset may degrade the performance of system depending on the coincidence window, and the effect
of event timer can be ignored. 
In addition, as we know that a 67 km
fiber-optical quantum key distribution system\,\cite{1367-2630-4-1-341} and
a 144 km free space transmission of entangled photon pairs have been
realized in experiments\,\cite{Ursin2007, Fedrizzi2009}, combining the two-photon assisted clock comparison scheme with the technologies in quantum communication, high precision
synchronization between long distance separated clocks could be realized in future.

\section*{Acknowledgments}
This work was supported by the National Natural Science Foundation of China under Grant Nos. 61475191, 61176084, and 11174282; and the Open Research Fund of Key Laboratory of Spectral Imaging Technology, Chinese Academy of Sciences.

\end{CJK*}
\end{document}